\PassOptionsToPackage{pdfpagelabels=false}{hyperref} 
\documentclass{acm_proc_article-sp}

\def\sharedaffiliation{%
\end{tabular}
\begin{tabular}{c}}
\usepackage{graphicx}
\begin{document}

\conferenceinfo{}{Bloomberg Data for Good Exchange 2016, NY, USA}

\title{Harnessing the Potential of the American Community Survey: Delving into Methods of Data Delivery}
\numberofauthors{2}
\author{
\alignauthor
Eve Ahearn\titlenote{Product Manager.}\\
    \email{eve.ahearn@enigma.io}
\alignauthor
Olga Ianiuk\titlenote{Data Engineer.}\\     
    \email{olga.ianiuk@enigma.io}
\sharedaffiliation
    \affaddr{Enigma Technologies Inc.}\\
    \email{245 Fifth Avenue}\\
    \email{New York, NY 10016}\\
    \email{+1 (800) 510-2856}\\
}

\maketitle

\begin{abstract}{The American Community Survey (ACS) is the bedrock underpinning any analysis of the US population, urban areas included. The Census Bureau delivers the ACS data in multiple formats, yet in each the raw data is difficult to export in bulk and difficult to sift through. We argue that Enigma's approach to the data delivery, such as our raw data and metadata presentation, reflects the survey's logical structure. It can be explored, interlinked, and searched; making it easier to retrieve the appropriate data applicable to a question at hand. We make the use of data more liquid via curated tables and API access; even metadata and notes from technical documentation are programmatically accessible. Additionally, we are working towards opening our scalable and reproducible ingestion process of ACS estimations. This paper details all of the ways the Census Bureau currently makes the data available, the barriers each of these raise to applying this data in analysis and how our approach overcomes them. Finally, this paper will address other recent innovations in making Census datasets more usable, the use cases suited to each and how they fit into the wider application of data science.}
\end{abstract}


\section{Introduction}
The Census Bureau publishes American Community Survey (ACS) data online in multiple different methods, alongside pages of accompanying documentation, and a dedicated data team to support it; yet, the data as delivered by the government presents high barriers to use. ACS contains demographic and socioeconomic details on the American population and the areas around them; samples are taken monthly and new data released each year. The data paints a portrait of the American populace, allowing for intelligent disbursement of government services. 

At Enigma, we have worked to make this critical dataset easier to browse, download in bulk and to apply to relevant problems. This paper introduces each of the ways that the Census Bureau distributes the data and the drawbacks of each, finally discussing Enigma's approach to delivery as well as that of other third parties. The methods through which datasets are published are of importance to data science, as the practice typically involves the use of data from external sources. By analyzing the data delivery landscape for a large and complex dataset such as ACS, we can understand how to better make critical data available for data scientists and how then to put this data to work for civic analysis. 

\section{Background on the ACS}
The ACS is an ongoing statistical survey covering a large and diverse spectrum of attributes about the United States' population including: income, housing stock, population mobility, country of origin, languages spoken, method of commute, health insurance status and many more. ACS serves two purposes: to provide continuous data--as opposed to the single snapshot captured by the Decennial Census--and to provide a greater level of detail on the United States' demographic, social, housing and economic characteristics. The Census Bureau sends out ACS surveys to one in 38 households every month and statistics from the survey are released every year, in one-year or five-year estimates. The status of the ACS program, only first implemented in 2005, is still in flux. Recent cuts in funding of the ACS have resulted in the 2015 cancellation of the three-year estimates, despite the widespread use of the data~\cite{_acs_2016}.

According to figures from the Census Bureau, results from the ACS guide the disbursement of \$400 billion in government funds and are also used by researchers, journalists, and public and private industry. Charitable organizations such as the Philadelphia-based Philabundance, use poverty data from the ACS to target aid~\cite{_communities_ACS}. Journalists at Bloomberg, for example, used data from the ACS to create an interactive visualization of which professions tended to marry each other~\cite{_who_marries_whom}, while companies such as Target~\cite{_target_2012}, 
analyze data from ACS to determine what merchanise to place at each store. Enigma itself used data from the ACS in a project called Smoke Signals, in partnership with the city of New Orleans and the American Red Cross, applying demographic data to predict which households might be without a smoke detector~\cite{_smoke_signals}.

Survey questions within the ACS can be very specific, spanning more than 60,000 variables in total. Examples of the topics covered by the housing subject, for example, include: the number of rooms in a given house, mortgage status by amount of real estate tax paid and type of kitchen a house contains. The specificity of these questions is what makes the survey so valuable for the disbursement of funds and the targeting of programs by governments and nonprofit organizations, but it also generates a large set of data that is difficult to apply and comprehend.

The ACS Summary File is divided into tabulated estimates for a certain geography, such as estimated number of people living in a specific county in Hawaii who speak Japanese at home, alongside a Margin of Error (MOE) for said estimation.\footnote{According to the 2014 5-year estimates, there are 38220 people living in Honolulu County, Hawaii, for instance, who speak Japanese at home --- with a margin of error of 1688 --- which is about 4\% of the island total population, see in \href{https://app.enigma.io/table/us.gov.census.acs.2014.5yr.language.language-spoken-at-home-by-ability-to-speak-english-for-the-population-5-years-and-over?search[]=\%40stusab\%20(\%22HI\%22)&search[]=\%40sumlevel\%20(\%22050\%22)&row=0&col=140&page=1&sort=b16001_069-}{language-spoken-at-home-by-ability-to-speak-english-for-the-population-5-years-and-over}}

The Summary File data is released yearly, in different estimations ranges. The ``2014 5-year" ACS data, for example, represents estimates collected for the five years between 2010 and 2014. The data in the 5-year survey represents greater geographic granularity than the ``2014 1-year" survey, which also includes only responses collected in the last year. In addition to the Summary File, the Census Bureau releases an anonymized version of the line by line survey responses called PUMS, or Public Use Microdata Sample. PUMS data can then be tabulated for cross-analyses not included in the Summary File, such as the median earnings of bike commuters. 

\section{Census Bureau's ACS Delivery}
A major area of focus in conversations around open data relates to ensuring released data is machine readable, unlike information published in formats such as PDF with embedded images.  Because of the complexity of ACS data, the Census Bureau has been extremely proactive in trying to make the data available through many different channels. The Census Bureau's site, American FactFinder\footnote{Available at \href{http://factfinder.census.gov/}{factfinder.census.gov}} provides a faceted search that enables users to locate data on specific geographies or topics. The Census also makes certain common estimates available via API\footnote{API stands for Application Program Interface. It is a standardized set of routines that enable one to programmatically request data from another machine. In this case request elements of the ACS Summary File database.} as well as offering raw data directly for download from the Census Bureau's FTP. The Census Bureau provides PDFs of  highly detailed technical documentation for each dataset and additionally publishes PDFs of errata notices, when it comes across errors in the data~\cite{_summary_2015}. The ACS-specific email help line has, in the experience of the authors, an impressive response time of fewer than 24 hours. The Census Bureau will even generate custom tabulations of the ACS data, for a cost of \$3,000 or more. There are further methods still of accessing ACS data from the Census Bureau --- they will load data onto a DVD and mail it if one was willing to pay and are unable to download large datasets. Yet, despite all of these efforts, ACS data as it is delivered by the source presents a number of obstacles. Each of these delivery methods, and the sacrifices each make between browsability and providing access to data in bulk, will be discussed in turn.

\subsection{Multiple Formats}
The American FactFinder (AFF) aims to make it easier to find popularly requested facts about a community, but the composition of its web interface makes it both difficult to discover the range of subjects for which information is available, and to access the underlying raw data\footnote{Note: The focus of this paper is on the Summary File and to that end, when the term ``raw data" is used it is referencing the Summary File data, not PUMS.} once a ``fact" of interest is identified.

AFF appears to be targeting a non data savvy or, at least, a non-ACS savvy audience. Little information about the structure of the survey is required. ``Prepackaged products", as the Bureau terms them, of data can be downloaded through the site, but a number of navigation pages must be passed through in order to be able to export data. As a consequence of this, the AFF doesn't allow users to access the data programmatically, but rather provides the means to drill down to a product or dataset of interest before a one-time export. 

It is possible for ACS data to be accessed programmatically either using the Census Bureau's FTP or through an API for a small subset of selected ACS data. However, both of these methods of delivery require the user to have significant familiarity with the survey to know which portions of the ACS are of interest. 

So while the API makes retrieving specific portions of the survey quite easy, due to the complexity of the survey it can be necessary to download the underlying raw data for the user to develop the knowledge needed to construct the appropriate API call. Therefore, outside of the AFF, users have to demonstrate meaningful data and technical literacy to utilize ACS data. The resulting user-friction keeps a lot of valuable and relevant ACS data out of the hands of those best positioned to use it.

\subsection{Summary File Organization}
The organization of the ACS Summary File data in the Census Bureau's FTP, which covers tabulated estimates, can be challenging to work with. The Summary Files are redundantly arranged in three different directories for each data release, each directory an attempt to address a different user need. This is done to account for a user's ability to work with \textit{very large files}. To this end, the source site  lists  an explicit warning that only experienced users are advised to work with directories that contain all of the data for numerous releases. 

The truth is that the file sizes are large enough to strain an internet connection, but are something a contemporary computer with some space to spare can handle. As of today, one needs less than 400GB to store all the expanded historical data locally, which is of help in analysis as the Census Bureau's FTP server can sometimes be sluggish.

Once any of the grouped files are obtained, one is expected to perform a number of transformations to get a subset of the desired columns and rows combinations. In particular, users are likely to adjoin the data with crucial parts of the metadata (e.g. schemas) that are provided in the ACS Summary File templates, and join data files that are published separately to explore integrated data from the tables. These transformations are roughly consistent across Summary File directories.

Assuming that Census Bureau controls for the data integrity among different Summary Files directories, those options are \textit{only} useful for users who intend to customize their Extract Transform Load (ETL) pipelines, while keeping the data synchronized with the FTP server.

\subsection{Excel Legacy}
The Census Bureau's delivery approach for the ACS Summary File data is optimized for opening each of the data and metadata files in spreadsheet software, in particular Microsoft's Excel, and in the software suite SAS. Consequently, a lot of the data delivery design solutions are based on an assumption that users prefer to use these tools as a primary ones to process the data. 

As SAS and Microsoft's Excel are proprietary software, users who want to access data in any other way have to deal with the plain text files (TXT), which were simply converted from Excel views (XLS), but not properly optimized for the change in data format. The issue here is that usually data intended for use in a spreadsheet software has cells merged to make their appearance more attractive to the human eye. This too, needs to be reverse engineered and is an obstacle encountered by users looking to apply the essential metadata enclosed in the ACS Sequence Table Number Lookup files. In general, Enigma argues that open formats must be used for open public data, otherwise perhaps open data files can be downloaded but not opened.

The notion of sequences is another consequence of the expectation of Excel. Sequence numbers were introduced to group ACS tables within similar subjects~\cite{_acs_2016}. The tables' assigned identifiers are composed from codified subjects and tables names and other attributes codes. In fact, all ACS tables can be aligned in a single giant spreedsheet, where columns indicate ACS questions and rows indicate geographical areas of a population sample being asked those questions. Columns are grouped into tables and tables are groubed by subjects. This logical organization is represented in the sequences numbers and while it reflects hierarchical structure of the data (see Section~\ref{we_del}), the delivery of the tables files in sequences governed by Excel limitations\footnote{Versions of Excel prior to version 12.0 limited a singular spreadsheet to 265 columns~\cite{_web_excel}.} is superficial (see  Section 2.3 in~\cite{_summary_2015}).

The Census Bureau provides instructions~\cite{_import_tool_2014} to compile the files, but this approach is hardly scalable, or is at least limited to the extent of a user's patience as they monotonously repeat Excel manipulations for different tables and states.

We suspect that one of the contributing factors in the use of sequence numbers was the large number of columns present in tables. It is true that it is hard to explore a table with a number of columns larger than a certain threshold. Nonetheless, even versions of Excel starting from version 12.0 support spreadsheets with a maximum of 16,384 columns~\cite{_web_excel}, while the largest ACS table at the moment contains only 526 columns\footnote{In Enigma these tables have 1052 columns as MOE columns are adjacent, e.g. \href{https://app.enigma.io/table/us.gov.census.acs.2014.5yr.industry-occupation-class-of-worker.class-of-worker-by-median-earnings-in-the-past-12-months-for-the-civilian-employed-population-16-years-and-over}{\scriptsize{industry-occupation-class-of-worker.class-of-worker-by-median-earnings-in-the-past-12-months}}} and the majority of tables (98.5\%) have less than 100 columns. For instance, the median of the number of columns for tables of the 2014 5-year release is 10 (the average is 58), at the same time 89\% of the tables have less than 50 columns, which in combination with adjacent MOE columns, results in table views of 100 columns at once.  

\section{Enigma's Efforts}
\label{we_del}
We use Enigma's own data pipeline tool ParseKit\footnote{For more information: \href{http://enigma.io/parsekit/}{enigma.io/parsekit}} to assemble different ACS parts. The tool has a Standard Library of ParseKit `steps' that was used for the basic Summary File directories handling and data and metadata files extraction. For the transformation part of the data processing pipeline a Custom Library of the ParseKit steps was written and we are working towards opening it in the future (see Section~\ref{future}). 

As a result of the ParseKit parser, a CSV file is produced for each table within data release requested. A subset of the targeted subjects can be supplied as well. CSV files for all tables, ACS releases and a complete file with the Census geographies descriptions for each release can be exported through Enigma's Public Data Explorer\footnote{Available at \href{https://app.enigma.io/search/source/us.gov.census.acs}{app.enigma.io/search/source/us.gov.census.acs}}. Full functionality of the ACS as accessed through the Public Data Explorer is discussed in Section \ref{browse_acs}.

The parsing process can be replicated and rerun with some frequency, as the Census Bureau will occasionally update older Summary Files if and when errors are discovered. Since the files need to be integrated under common schemas, as a parser byproduct we create a local database of thematic tables. This is not a necessary part of the parsing process, but in doing so the database can then be used as a structure with which to query ACS data. Besides the fundamental data and metadata merging implemented with the aid of the database, we link each record with its associated Census geography unit by default providing an area name, logical identifier, state abbreviation and geographic summary level, although these attributes can be customized. Finally, we link each estimation point with its statistical random sampling error metric, MOE, precalculated by the Census Bureau.

Using the approach described above, Enigma has parsed all Summary File data available on the FTP. The infrastructure advantages allow us to control for the potential data inconsistencies and incompleteness. We have not yet found any serious problems with the raw data quality, although we did report a few missing geographies to the source.

\newpage
\subsection{Browsing the ACS} \label{browse_acs}
ACS data is accessible through Enigma's web application and data search engine, Public Data Explorer. Data can be browsed through or searched by topic, enabling the user to explore the vast quantity of ACS estimates. Once an item of interest is identified, the web platform allows the user to perform quick statistics, such as calculating the median or standard deviation for a particular response, or to identify the frequency at which data for certain geographic levels is available. 

The ACS data and metadata can additionally be browsed by subject, sorted by geography and the table descriptions can be keyword searched. The latter especially is helpful to check for an imputed data corresponding to a certain estimation or MOE. This enables a user to engage with and get a feel for the data prior to downloading it locally. Once a user has located a table--or set of estimates--of interest, they can be accessed via the web application's API or exported data from the site. 

Data in Enigma is cataloged by the source of the data and topical tags and the ACS data is presented as a set of hierarchically dependent relationships that we found natural to follow. Within Enigma, each table can then be uniquely identified by its survey release (year and period), an ACS thematic subject and by a human readable name of a thematic table, for example: \href{https://app.enigma.io/table/us.gov.census.acs.2014.5yr.age-sex.median-age-by-sex}{age-sex.median-age-by-sex}. The subjects' and tables' logical identifiers assigned by the Census Bureau can be accessed either through the metadata or the columns field names within those tables. On the contrary, columns' logical identifiers are the actual field names, while each column has a human-readable display name that can be viewed in the web application and queried by its metadata. For the table above, the table identifier B01002, subject ID `01' stands for Age and Sex and table ID `002' for Median Age by Sex, is attached to columns names, e.g. b01002\_003, b01002\_003\_moe that are fully described as Female Percentage, `003', and Margin of Error for Percentage Female, `003\_moe'. 

Data for all states in a particular topical category is kept together in a singular table intentionally as to facilitate a bulk data export. Users can then browse the data to explore the range of geographic specificity for which data for a particular topic is published, from census tract to city, up to statewide estimates. Browsing is especially helpful in the case of geography as estimates are released for varying degrees of geographic granularity depending on the sensitivity of the subject.

\subsection{Curation}
In any re-representation of data there is a conflict between an interest in remaining true to the source and a desire to correct any errors or ambiguities within the data. In our representation of ACS, we change nothing of the data itself, wishing to remain as true to source as possible. However, our views of the data are curatorial in the sense that we present the survey responses in a table already joined with the names of the geographic locations and the MOE. This is done as we anticipate that any use of the data will necessitate looking up what geographic IDs such as ``040C0US02" represent. The value of the MOE is discussed below.

In its distribution of the data the Census Bureau makes the complexity of the information clear, and in our republishing of the data we have tried to retain that by preserving the MOE available for each variable. As is inherent in surveys, the MOE is larger the smaller the sample size, and any user of the ACS should be aware that the MOE for a small geographic area will be substantial. For instance, the 2014 5-year ACS can tell us that approximately 60 people living within Census Tract 707 boundaries in Carter County, Tennessee, commute to work by carpool --- with a MOE of 48. In other words, if we were to conduct 100 different polls on the same-size samples drawn from this area, we would expect the answers in 90 of those polls to be within 48 units from the true count of the residents carpooling from Tract 707, Tennessee to work\footnote{This estimate should be considered as a highly unreliable with the $48.6\%$ coefficient of the variation~\cite{_esri_2014}, see in \href{https://app.enigma.io/table/us.gov.census.acs.2014.5yr.journey-to-work.means-of-transportation-to-work?row=0&col=6&page=1}{\scriptsize{journey-to-work.means-of-transportation-to-work}}} or there are between 12 and 108 `carpoolers'. This uncertainty is difficult to summarize with a chart.

We coerce all the estimations and MOE values to a numeric type. This allows us to provide functionality calculating basic descriptive statistics for each of the survey questions. However, as a drawback, this necessitates treating missing values in a different manner from that of the Census Bureau. It is crucial to evaluate and report the data uncertainty for a quality research. As a future improvement, we plan to apply our typical technique and offset all the special values to adjacent columns and provide a comprehensive description of the missing values in the metadata. This is also another step of making information about the imputed values more accessible.

\section{Other Efforts}
The authors are not the only ones interested in facilitating a better delivery of the information contained in the ACS. Other projects have arisen recently for much the same aim, albeit taking a different approach and with different audiences in mind. 

In 2016, Deloitte, Datawheel and the MacroConnections project out of the MIT Media Lab launched Data USA\footnote{Available at \href{http://datausa.io/}{datausa.io}}, terming it the ``most comprehensive visualization of U.S. Public Data." Much of the data on the site comes from the ACS. The site has pre-made visualizations that function as ``profiles" of given geographies, industries or occupations. Data from any one of the visualizations can be downloaded, but the intention of the site appears to be to demonstrate what can be gleaned from the data, and to make the data available to those who might not otherwise create their own visualizations. 
The ideal audience of the tool then, is unlikely to be a data scientist. Similar to Data USA, Social Explorer\footnote{Available at \href{http://www.socialexplorer.com/}{socialexplorer.com}}, a demographic data company, emphasizes visualization of the ACS.

Census Reporter\footnote{Available at \href{http://censusreporter.org/}{censusreporter.org}} is a Knight Foundation funded project that aims to make census data easier for journalists to write stories using information from census surveys. The Census Reporter has its own API with access to ACS data, though the construction of an API call still requires knowledge of the needed table ID and geographic ID. The Census Reporter project provides quite helpful explanations of the structure, after a close reading of which, a user can determine what are the tables and geographies of interest. The site also has sample scripts and instructions to guide users on getting their own local copy of the ACS data. However, the project does not attempt a visual representation of the entire data release. Enigma's approach to this ``browsability" of the data is discussed in Section~\ref{we_del}.

The number of projects aiming to make ACS easier to use is no accident, the information in the survey is critical, and very helpful for data science projects of all kinds. However, the data is inherently complex, and is valuable due to its complexity, so there is a limit to how simple the delivery of the data can be. There is rigorous social science~\cite{nancy_torrieri2014} and a number of nuances~\cite{_polls_1997} behind the construction of the survey and a user needs to be data literate enough to apply the insights with caution, keeping the MOE, for example, in mind. As an additional complication, the Census Bureau notes that estimates from Summary Files years with non-overlapping periods cannot necessarily be compared directly, as the phrasing of the question might have changed or obtained samples are not statistically different~\cite{_ofm_acs_user_guide}.

While the actors described above all seek to make bulk data from the ACS more accessible or digestible, there are also innumerable journalistic and policy projects incorporating information from the ACS.

\section{Better ACS Delivery}
\label{better}
One of Enigma's primary missions is to make public datasets of critical importance more accessible. We are very thankful to those who create and publish high quality public data. This paper makes it clear that the Census Bureau in general and the ACS section in particular is an example of such a data source. Our aim is to be useful in return and offer our expertise and infrastructure to unleash the maximum potential of the ACS.

In one scenario of an idealized state of delivery: the usage of the sequences would be suspended and the data from thematic tables instead grouped by the ACS thematic subjects and ACS thematic universes; the data files names would contain table identifiers and  the tables grouping archives names would contain subjects identifiers. As there are no subjects that comprise more than 150 tables, no individual group archive will contain significant number of data files and so selection of the tables of interest is more straightforward. All the data files should contain table identifiers and column identifiers in their headers. This way, it would be painless to provide any additional metadata in a hierarchically structured data dictionary with subjects, tables and columns names as the dictionary keys. All of the crucial metadata should be provided within that single dictionary. 

We think that the Census Bureau could upgrade their  approach to the delivery of the ACS for data science with a consideration of our outline of the data delivery design drawn from our own data munging experience, the way we re-deliver the data and other observations described in this paper.

\subsection{Future Work}
\label{future}
There is future work to be done both in the study of optimized methods of data delivery and in our specific delivery of the ACS. For the former, future work in this area could focus on data delivery of other sources of datasets or could involve user research on data acquisition methods and preferences. This paper speaks from the authors' specific experience; we expect other audiences to have other expectations and preferences in terms of accessing and understanding data. This is an ever-evolving topic. As new technologies for publishing and analyzing data emerge, preferences for data delivery are likely to change along with them. 

In the case of Enigma's representation of the ACS we have a number of specific goals to further improve our method of delivery. We plan to deepen our curation of the data, further reducing the friction on analysis of a singular geography across subjects. We also plan to experiment further with adding additional information from the technical documentation into the structured metadata attached to the datasets. In particular, adding in greater context around nulls in the data so that any analysis using the data can take the manner of uncertainty into account.
    
Beyond improving our data ingestion practices, we are also working to make these processes themselves more transparent. ParseKit, which started as an internal tool, is now being used by a number of external commercial users as well as growing number of governmental and nonprofit actors such as the California Water Authority and the City of New Orleans.

\subsection{Conclusion}
The authors' approach to the delivery of ACS attempts to allow a user to obtain critical demographic data in a manner amenable to data science while avoiding the need to study technical documentation at length, download an entire survey year in bulk and handle peculiarities within the raw data. Our delivery method optimizes for an understanding of the data at hand, with documentation, the Summary File data and metadata smoothly linked. Information contained within the ACS is critical to understanding and improving American cities and yet before the information can be put to use it must be truly accesible.

\section{Acknowledgments}
The authors thank our coworkers at Enigma for [many] discussions on data delivery in general and the American Community Survey in particular.

\nocite{*}
\bibliographystyle{abbrv}
\bibliography{ACS_paper}

\end{document}